\documentstyle[12pt,aaspp]{article}
\slugcomment{Preprint IAS-AST-93/70 and MIT-CSR-93-32, subm. ApJ}

\def\ltsima{$\; \buildrel < \over \sim \;$}
\def\lsim{\lower.5ex\hbox{\ltsima}}
\def\gtsima{$\; \buildrel > \over \sim \;$}
\def\gsim{\lower.5ex\hbox{\gtsima}}
\begin{document}
\title{Second Order Power Spectrum and Nonlinear Evolution at High Redshift}
\author{Bhuvnesh Jain and Edmund Bertschinger\altaffilmark{1}}
\affil{Department of Physics, MIT, Cambridge, MA 02139 USA}
\altaffiltext{1}{Also Institute for Advanced Study, Princeton, NJ 08540}

\begin{abstract}
The Eulerian cosmological fluid equations are used to study the nonlinear
mode coupling of density fluctuations.
We evaluate the second-order power spectrum including all four-point
contributions.
In the weakly nonlinear regime we find that the dominant nonlinear
contribution for realistic cosmological spectra is made by the coupling
of long-wave modes and is well estimated by second order perturbation
theory.
For a linear spectrum like that of the cold dark matter model,
second order effects cause a significant
enhancement of the high $k$ part of the spectrum and a slight suppression
at low $k$ near the peak of the spectrum.
Our perturbative results agree well in the quasilinear regime with the
nonlinear spectrum from high-resolution N-body simulations.

We find that due to the long-wave mode coupling, characteristic nonlinear
masses grow less slowly in time (i.e., are larger at higher redshifts) than
would be estimated using the linear power spectrum.
For the cold dark matter model at $(1+z)=(20,10,5,2)$ the
nonlinear mass is about $(180,8,2.5,1.6)$ times (respectively) larger
than a linear extrapolation would indicate, if the condition rms $\delta
\rho/\rho =1$ is used to define the nonlinear scale.  At high redshift
the Press-Schechter mass distribution significantly underestimates the
abundance of high-mass objects for the cold dark matter model.
Although the quantitative results depend on the
definition of the nonlinear scale, these basic
consequences hold for any initial spectrum whose post-recombination
spectral index $n$ decreases sufficiently rapidly with increasing $k$,
a feature which arises quite generally during the transition from a
radiation- to matter-dominated universe.
\end{abstract}

\keywords{cosmology: theory --- large-scale structure of universe
--- galaxies: clustering --- galaxies: formation}

\section{Introduction}

There exists a standard paradigm for the formation of cosmic structure:
gravitational instability in an expanding universe.  According to this
paradigm, dark matter
density fluctuations $\delta(\vec x\,)\equiv\delta\rho(\vec x\,)/
\bar\rho$ created in the early universe lay dormant until the universe
became matter-dominated at a redshift $z=z_{\rm eq}\approx 2.5\times10^4\,
\Omega h^2$ (where $\Omega$ is the present density parameter for
nonrelativistic matter and the present Hubble parameter is $H_0=100\,h\
{\rm km\, s^{-1}}{\,\rm Mpc}^{-1}$).  After
this time, the density fluctuations increased in amplitude as predicted by
the well-known results of linear perturbation theory (e.g., Peebles 1980;
Efstathiou 1990; Bertschinger 1992), until the fluctuations became nonlinear
on some length scale.  Bound condensations of this scale then collapsed and
virialized, forming the first generation of objects (Gunn \& Gott 1972;
Press \& Schechter 1974).  Structure formation then proceeded hierarchically
as density fluctuations became nonlinear on successively larger scales.

At early times density fluctuations were small on the length scales
of present day large-scale structure. Therefore, after the universe
became matter dominated fluctuations on scales much larger than
the scale of collapsed objects can be studied
under the approximation of a pressureless, irrotational fluid evolving under
the action of Newtonian gravity. A perturbative analysis of the fluid
equations in Fourier space can then be used to study the effects of mode
coupling between scales that are weakly nonlinear. This is the approach
we shall follow in this paper. Nonlinear analyses in real and Fourier space
are somewhat complementary in that real space analyses are best suited
to studying the effect of nonlinearities on the collapse and shapes
of individual objects (Bertschinger \& Jain 1993), whereas Fourier space
studies provide estimates
of how different parts of the initial spectrum couple and influence the
evolution of statistical quantities like the power spectrum. In principle
of course, the two approaches are equivalent and should give the same
information. For perturbative analyses in real space see,
e.g., Peebles (1980), Fry (1984), Hoffman (1987), Zaroubi \& Hoffman (1993),
and references therein.

Although density fluctuations of different wavelengths evolve
independently in linear perturbation theory, higher order
calculations provide an estimate of some  nonlinear effects.
Preliminary second order analyses have led to the conventional view that
in models with decreasing amounts of power on larger scales
long-wavelength fluctuations have no significant effect on the
gravitational instability occuring on small scales.
On the other hand, it is known that under some circumstances
small-scale, nonlinear waves can transfer significant amounts of power
to long-wavelength, linear waves. If the initial spectrum
is steeper than $k^4$ at small $k$ (comoving wavenumber), then small-scale,
nonlinear waves can transfer power to long wavelength linear waves
so as to produce a $k^4$ tail in the spectrum.
(Zel'dovich 1965; Peebles 1980, Section 28;
Vishniac 1983; Shandarin \& Melott 1990).

The question of whether power can be transfered from large to small scales was
examined by Juszkiewicz (1981), Vishniac (1983), Juszkiewicz, Sonoda \& Barrow
(1984),  and more recently by Coles (1990), Suto and Sasaki (1991)
and Makino, Sasaki and Suto (1992). Their analyses involved writing down
integral expressions for the second order contribution to the
power spectrum, examining their limiting forms and evaluating
them for some forms of the initial spectrum. Juszkiewicz et al. (1984)
examined the autocorrelation function and found that the clustering
length decreases due to power transfer from large to small
scales for the initial spectrum $P(k) \propto k^2$. However, for the
cold dark matter (hereafter CDM) spectrum Coles (1990) found the
opposite effect, though it is not significant unless $\sigma_8$ is
taken larger than 1.
Makino et al. (1992) have analytically obtained the second order
contributions for power law spectra, and estimated the contribution
for the CDM spectrum by approximating it as two power laws.
Bond \& Couchman (1988) have compared the second order CDM power spectrum
to the Zel'dovich approximation evaluated at the same order.
Some issues of mode coupling have recently been investigated
through N-body simulations in $2$-dimensions (see e.g.,
Beacom et al. 1991; Ryden \& Gramann 1991; Gramann 1992).

We have used the formalism developed in some of the perturbative
studies cited above, and especially by Goroff et al. (1986), to calculate
second order contributions to the power spectrum (i.e., up to fourth order
in the initial density) for the standard CDM spectrum.
Second order perturbation
theory has a restricted regime of validity, because once the density
fluctuations become sufficiently large the perturbative expansion
breaks down. For this reason N-body simulations have been used more
extensively to study the fully nonlinear evolution of density fluctuations.
However, perturbation theory is very well suited to address some specific
aspects of nonlinear evolution and to provide a better understanding
of the physical processes involved. Being less costly and time-consuming
than N-body simulations, it lends itself easily to the study of different
models.  Perturbation theory should be considered a complementary technique
to N-body simulations, for while its validity is limited, it does not suffer
from the resolution limits that can affect the latter. Hence by comparing
the two techniques their domains of validity can be tested and their
drawbacks can be better understood. In this paper we shall make
such comparisons for the CDM spectrum.

The most powerful use of
perturbative calculations is in the study of weakly nonlinear evolution
out to very high redshifts, spanning decades of comoving
length scales in the spectrum. Since the formulation of the perturbative
expansion allows for the time evolution of the spectrum to be obtained
straightforwardly, we obtain the scaling in time of characteristic
nonlinear mass scales ranging from the nonlinear scale today, about
$10^{14} \,M_\odot$, to about $10^5 \,M_\odot$, the smallest baryonic mass
scale likely to have gone nonlinear after the universe became matter
dominated. Such an analysis cannot be done by existing N-body simulations
as the dynamic range required to cover the full range of scales
with adequate spectral resolution exceeds that of the current
state-of-the-art.

There are two principal limitations to our analytic treatment:
the first arises from
the general problem that the perturbative expansion breaks down when
nonlinear effects become sufficiently strong.
This drawback is particularly severe in our case because the regime of
validity is difficult to estimate. It is reasonable to expect that
second order perturbation theory ceases to be valid when the rms
$\delta \rho/\rho \gsim 1$, but one cannot be more precise without explicitly
calculating higher order contributions.

The second kind of limitation arises from the simplifying assumptions that
pressure and vorticity are negligible. On small enough
scales nonlinear evolution causes the intersection of particle orbits
and thus generates pressure and vorticity. Through these effects
virialization on small-scales can alter the growth of fluctuations on
larger scales.
It is plausible that the scales in the weakly nonlinear regime are
large enough that this effect is not significant.
This belief is supported by heuristic arguments as well as recent studies of
N-body simulations (Little, Weinberg
\& Park 1991; Evrard \& Crone 1992 and references therein).
We conclude that the first kind of limitation, namely the neglect of
higher order contributions, or worse still, the complete breakdown of the
perturbative expansion, is likely to be more severe for our results.
We shall address this where appropriate and accordingly attempt to draw
conservative conclusions supported by our own N-body simulations.

The formalism for the perturbative calculation is described in Section
2. We describe the numerical results for CDM in Section 3.1 and
compare them to N-body simulations in Section 3.2. The scaling of the
nonlinear scale as a function of redshift is presented in Section 3.3.
The distribution of nonlinear masses is examined in Section 3.4
We discuss cosmological implications of the results in Section 4.

\section{Perturbation Theory}
In this section we describe the formalism for perturbative solutions
of the cosmological fluid equations in Fourier space. Our approach
is similar to that of Goroff et al. (1986). The formal perturbative
solutions are then used to write down the explicit form of the second order
contribution to the power spectrum.

\subsection{General Formalism}

We suppose for simplicity that the matter distribution after recombination
may be approximated as a pressureless fluid with no vorticity.
We further assume that peculiar velocities are nonrelativistic and that the
wavelengths of interest are much smaller than the Hubble distance $cH^{-1}$
so that a nonrelativistic Newtonian treatment is valid.  Using comoving
coordinates $\vec x$ and conformal time $d\tau=dt/a(t)$, where $a(t)$ is the
expansion scale factor, the nonrelativistic cosmological fluid equations are
\begin{mathletters}
\begin{equation}
{\partial\delta\over\partial\tau}+\vec\nabla\cdot[(1+\delta)\vec v\,]=0 \ ,
\label{continuity}
\end{equation}
\begin{equation}
{\partial\vec v\over\partial\tau}+\left(\vec v\cdot
\vec\nabla\right)\vec v=-{\dot a\over a}\,\vec v-\vec\nabla\phi\ ,
\label{euler}
\end{equation}
\begin{equation}
\nabla^2\phi=4\pi Ga^2\bar\rho\delta\ ,
\label{poisson}
\end{equation}
where $\dot a\equiv da/d\tau$.
Note that $\vec v\equiv d\vec x/d\tau$ is the proper peculiar velocity,
which we take to be a potential field so that $\vec v$ is fully
specified by its divergence:
\begin{equation}\theta\equiv\vec\nabla\cdot\vec v\ .
\label{u}\end{equation}
\end{mathletters}
We assume an Einstein-de Sitter ($\Omega=1$) universe, with $a\propto
t^{2/3}\propto\tau^2$. We will also assume that the initial
(linear) density fluctuation field is a gaussian random field.

To quantify the amplitude of fluctuations of various scales it is preferable
to work with the Fourier transform of the density fluctuation field,
which we define as
\begin{equation}\hat\delta(\vec k,\tau)=\int{d^3x\over(2\pi)^3}\,
e^{-i\vec k\cdot\vec x}\,\delta(\vec x,\tau)\ ,
\label{deltak}\end{equation}
and similarly for $\hat\theta(\vec k,\tau)$.  The power spectrum
(power spectral density) of $\delta(\vec x,\tau)$ is defined by
the ensemble average two-point function,
\begin{equation}\langle\hat\delta(\vec k_1,\tau)\,
\hat\delta(\vec k_2,\tau)\rangle=
P(k_1,\tau)\,\delta_{\rm D}(\vec k_1+\vec k_2)\ ,
\label{pk}\end{equation}
where $\delta_{\rm D}$ is the Dirac delta function, required for a spatially
homogeneous random density field.  For a homogeneous and isotropic random
field the power spectrum depends only on the magnitude of the wavevector.
The contribution to the variance of $\delta(\vec x,\tau)$ from waves in the
wavevector volume element $d^3k$ is $P(k,\tau)d^3k$.

Fourier transforming equations (1) gives
\begin{mathletters}
\begin{equation}{\partial\hat\delta\over\partial\tau}+\hat\theta=
-\int\!d^3k_1\!\int\!d^3k_2\,\delta_{\rm D}(\vec k_1+\vec k_2-\vec k\,)
\,{\vec k\cdot\vec k_1\over k_1^2}\,
\hat\theta(\vec k_1,\tau)\,\hat\delta(\vec k_2,\tau)\ ,
\label{fourier1}\end{equation}
\begin{equation}{\partial\hat\theta\over\partial\tau}+{\dot a\over a}\,
\hat\theta+{6\over\tau^2}\,\hat\delta=
-\int\!d^3k_1\!\int\!d^3k_2\,\delta_{\rm D}(\vec k_1+\vec k_2-\vec k\,)
\,{k^2(\vec k_1\cdot\vec k_2)\over2k_1^2 k_2^2}\,
\hat\theta(\vec k_1,\tau)\,\hat\theta(\vec k_2,\tau)\ .
\label{fourier2}\end{equation}
\end{mathletters}

In equations (4) the nonlinear terms constitute the right-hand
side and illustrate that the nonlinear evolution of the fields
$\hat\delta$ and $\hat\theta$ at a given wavevector $\vec k$ is determined by
the mode coupling of the fields at all pairs of wavevectors
whose sum is $\vec k$, as required by spatial homogeneity. This makes
it impossible to obtain exact solutions to the equations, so that the
only general analytical technique for self-consistently evaluating the
nonlinear terms is to make a perturbative expansion in $\hat\delta$ and
$\hat\theta$. The formalism for such an expansion has been systematically
developed by Goroff et al. (1986) and recently
extended by Makino et al. (1992). Following these authors we
 write the solution to equations (4) as a perturbation series,
\begin{equation}
\hat\delta(\vec k,\tau)=\sum_{n=1}^\infty a^n(\tau)\,\delta_n(\vec k\,)
\ ,\quad\hat\theta(\vec k,\tau)=\sum_{n=1}^\infty\dot a(\tau)a^{n-1}(\tau)
\,\theta_n(\vec k\,)
\label{pert}\end{equation}
It is easy to verify that for $n=1$ the time dependent part of the
solution correctly gives the
linear growing modes $\delta_1 \propto a(\tau)$ and $\theta_1
\propto \dot a $ and that the time-dependence is consistent with
equations (4) for all $n$.
To obtain formal solutions for the $\vec k$ dependence at all orders we
proceed as follows.

Substituting equation (5) into equations (4) yields, for $n>1$,
\begin{equation}
n\delta_n(\vec k\,)+\theta_n(\vec k\,)=A_n(\vec k\,)\ ,\quad 3\delta_n(\vec
k\,)+(1+2n)\theta_n(\vec k\,)=B_n(\vec k\,)\ ,
\label{recurs0}\end{equation}
where
\begin{mathletters}
\begin{equation}A_n(\vec k\,)\equiv
-\int\!d^3k_1\!\int\!d^3k_2\,\delta_{\rm D}(\vec k_1+\vec k_2-\vec k\,)
\,{\vec k\cdot\vec k_1\over k_1^2}\sum_{m=1}^{n-1}
\theta_m(\vec k_1)\,\delta_{n-m}(\vec k_2)\ ,
\label{recurs1}\end{equation}
\begin{equation}
B_n(\vec k\,)\equiv-\int\!d^3k_1\!\int\!d^3k_2\,\delta_{\rm D}
(\vec k_1+\vec k_2-\vec k\,)
\,{k^2(\vec k_1\cdot\vec k_2)\over k_1^2 k_2^2}\sum_{m=1}^{n-1}
\theta_m(\vec k_1)\,\theta_{n-m}(\vec k_2)\ .
\label{recurs2}\end{equation}
\end{mathletters}
Solving equations (6) for $\delta_n$ and $\theta_n$ gives, for $n>1$,
\begin{equation}
\delta_n(\vec k\,)={(1+2n)A_n(\vec k\,)-B_n(\vec k\,)\over(2n+3)(n-1)}\ ,
\quad\theta_n(\vec k\,)={-3A_n(\vec k\,)+nB_n(\vec k\,)\over(2n+3)(n-1)}
\ .
\label{recurs3}\end{equation}

Equations (7) and (8) give recursion relations for $\delta_n(\vec k\,)$
and $\theta_n(\vec k\,)$, with starting values $\delta_1(\vec k\,)$ and
$\theta_1=-\delta_1$.  The general solution may be written
\begin{mathletters}
\begin{equation}
\delta_n(\vec k\,)=\int d^3q_1\cdots\int d^3q_n\,
\delta_{\rm D}(\vec q_1+\cdots+\vec q_n-\vec k\,)\,
F_n(\vec q_1,\ldots,\vec q_n)\,
\delta_1(\vec q_1)\cdots\delta_1(\vec q_n)\ ,
\label{deltan}\end{equation}
\begin{equation}
\theta_n(\vec k\,)=-\int d^3q_1\cdots\int d^3q_n\,
\delta_{\rm D}(\vec q_1+\cdots+\vec q_n-\vec k\,)\,
G_n(\vec q_1,\ldots,\vec q_n)\,
\delta_1(\vec q_1)\cdots\delta_1(\vec q_n)\ .
\label{un}\end{equation}
\end{mathletters}
 From equations (7)--(9) we obtain recursion relations for $F_n$ and $G_n$:
\begin{mathletters}
\begin{eqnarray}
F_n(\vec q_1,\ldots,\vec q_n)&=&\sum_{m=1}^{n-1}
{G_m(\vec q_1,\ldots,\vec q_m)\over(2n+3)(n-1)}
\Biggl[(1+2n){\vec k\cdot\vec k_1\over k_1^2}
F_{n-m}(\vec q_{m+1},\ldots,\vec q_n)\nonumber\\
&&+{k^2(\vec k_1\cdot\vec k_2)\over
k_1^2k_2^2}G_{n-m}(\vec q_{m+1},\ldots,\vec q_n)\Biggr]\ ,
\label{fn}\end{eqnarray}
\begin{eqnarray}
G_n(\vec q_1,\ldots,\vec q_n)&=&\sum_{m=1}^{n-1}
{G_m(\vec q_1,\ldots,\vec q_m)\over(2n+3)(n-1)}
\Biggl[3{\vec k\cdot\vec k_1\over k_1^2}
F_{n-m}(\vec q_{m+1},\ldots,\vec q_n)\nonumber\\
&&+n{k^2(\vec k_1\cdot\vec k_2)\over
k_1^2k_2^2}G_{n-m}(\vec q_{m+1},\ldots,\vec q_n)\Biggr]\ ,
\label{gn}\end{eqnarray}
\end{mathletters}
where $\vec k_1\equiv\vec q_1+\cdots+\vec q_m$, $\vec k_2\equiv\vec q_{m+1}
+\cdots+\vec q_n$, $\vec k\equiv\vec k_1+\vec k_2$ and $F_1=G_1=1$.
Equations (10) are equivalent to equations (6) and (A1) of
Goroff et al. (1986), with $F_n=P_n$ and $G_n=(3/2)Q_n$ in their notation.

\subsection{Power Spectrum at Second Order}
To calculate the power spectrum we shall prefer to use
symmetrized forms of $F_n$ and $G_n$, denoted $F_n^{(s)}$ and
$G_n^{(s)}$ and obtained by summing the $n!$ permutations of $F_n$ and
$G_n$ over their $n$ arguments and dividing by $n!$. Since the arguments are
dummy variables of integration the symmetrized functions can
be used in equations (9) without changing the result.
The symmetrized second-order solutions of equations (10) are given by
\begin{mathletters}
\begin{equation}
F_2^{(s)}(\vec k_1,\vec k_2)={5\over7}+{2\over7}{(\vec k_1\cdot\vec
k_2)^2\over k_1^2k_2^2}+{(\vec k_1\cdot\vec k_2)\over2}\left({1\over
k_1^2}+{1\over k_2^2}\right)\ ,
\label{f2}\end{equation}
\begin{equation}
G_2^{(s)}(\vec k_1,\vec k_2)={3\over7}+{4\over7}
{(\vec k_1\cdot\vec k_2)^2\over k_1^2k_2^2}+{(\vec k_1\cdot\vec k_2)\over2}
\left({1\over k_1^2}+{1\over k_2^2}\right)\ .
\label{g2}\end{equation}
\end{mathletters}
Note that $F^{(s)}_2$ and $G^{(s)}_2$ have first-order poles as $k_1\to0$
or $k_2\to0$ for fixed $\vec k$: $F^{(s)}_2\sim G^{(s)}_2\sim(1/2)
\cos\vartheta\,(k_1/k_2+k_2/k_1)$ where $\vartheta$ is the angle between
$\vec k_1$ and $\vec k_2$. The expression for $F_3^{(s)}$ will also
be required, but since it is very long we shall wait to write a
simplified form below.

The recursion relations in equations (10) may be used
to compute the power spectrum at any order
in perturbation theory.  Substituting equation (5) into equation (3), we have
\begin{eqnarray}
P(k,\tau) \,\delta_{\rm D}(\vec k+\vec k')&=&
\langle\delta(\vec k,\tau)\,\delta(\vec k',\tau)\rangle \nonumber\\
&=&a^2(\tau)\langle\delta_1(\vec k)\,\delta_1(\vec k')\rangle+
a^4(\tau)\Biggl[\langle\delta_1(\vec k)\,\delta_3(\vec k')\rangle+
\langle\delta_2(\vec k)\,\delta_2(\vec k')\rangle \nonumber\\
&&+\langle\delta_3(\vec k)\,\delta_1(\vec k')\rangle\Biggr]+ O(\delta_1^6)\ .
\label{pkpert}\end{eqnarray}
Equation (12) explicitly shows all the terms contributing to the power
spectrum at fourth order in the initial density field $\delta_1$
(or second order in the initial spectrum), as the $n$th order field
$\delta_n(\vec k)$ involves $n$ powers of $\delta_1(\vec k)$.
With the definition
\begin{equation}
\langle\delta_m(\vec k)\,\delta_{n-m}(\vec k')\rangle\equiv
P_{m,n-m}(k)\,\delta_{\rm D}(\vec k+\vec k')\ 
\label{pmn}\end{equation}
the power spectrum up to second order (i.e., fourth order in $\delta_1$)
is given by equation (12) as
\begin{eqnarray}
P(k, \tau)&=&a^2(\tau)P_{11}(k)+a^4(\tau)[P_{22}(k)+2 P_{13}(k)]\nonumber\\
&=&a^2(\tau)P_{11}(k)+a^4(\tau)P_2(k) \, ,
\label{pnet1}\end{eqnarray}
where the net second order contribution $P_2(k)$ is defined as
\begin{equation}
P_2(k) = P_{22}(k)+2 P_{13}(k)\, .
\label{p2}\end{equation}

To determine $P_2(k)$ we need to evaluate the $4$-point
correlations of the linear density field $\delta_1(\vec k\,)$.
For a
gaussian random field, all cumulants (irreducible correlation functions)
of $\delta_1(\vec k\,)$ vanish aside from the 2-point cumulant, which is
given by equation (3) for $m=n-m=1$.  All odd moments of $\delta_1(\vec
k\,)$ vanish.  Even moments are given by symmetrized products of the
2-point cumulants.  Thus the $4$-point correlation function
of $\delta_1(\vec k)$ is
\begin{eqnarray}
&&\langle\delta_1(\vec k_1)\,\delta_1(\vec k_2)\,\delta_1(\vec k_3)\,
\delta_1(\vec k_4)\rangle=P(k_1)P(k_3)\delta_{\rm D}(\vec k_1+\vec k_2)
\delta_{\rm D}(\vec k_3+\vec k_4)\nonumber\\
&&+P(k_1)P(k_2)\delta_{\rm D}(\vec k_1+\vec k_3)\delta_{\rm D}(\vec k_2
+\vec k_4)+P(k_1)P(k_2)\delta_{\rm D}(\vec k_1+\vec k_4)\delta_{\rm D}
(\vec k_2+\vec k_3)\ .
\label{gauss}\end{eqnarray}

With the results and techniques described above, we can proceed to
obtain the second order contribution to the power spectrum. The two
terms contributing at second order simplify to the following $3$-dimensional
integrals in wavevector space:
\begin{equation}
P_{22}(k)=2 \int\!d^3q\,P_{11}(q)\,P_{11}(|\vec k-\vec q|)
\left[F_2^{(s)}(\vec q,\vec k-\vec q)\right]^2 \ , 
\label{p22}\end{equation}
with $F_2^{(s)}$ given by equation (11a), and
\begin{equation}
2 P_{13}(k)=6P_{11}(k)\!\int\!d^3q\,P_{11}(q)\,
F_3^{(s)}(\vec q,-\vec q,\vec k\,)\ .
\label{p13a}\end{equation}
The numbers in front of the integrals arise from the procedure
of taking expectation values illustrated in equation (16).
We write the integrals in spherical coordinates $q, \vartheta$, and $\phi$:
the magnitude, polar angle and azimuthal angle, respectively,
 of the wavevector $ \vec q $. Then with the external wavevector $\vec k$
aligned along the $z$-axis
the integral over $\phi$ is trivial and simplifies $\int d^3q$ to
$2\pi\int dq \, q^2\int d\cos\vartheta$. For $P_{13}$, the dependence on
$\vartheta$ is also straightforward as it arises only through $F_3^{(s)}$
and not $P_{11}$. This allows the integral over $\cos\vartheta$ to be done
analytically as well, giving (Makino et al. 1992)
\begin{eqnarray}
2 P_{13}(k)&=&{2\pi\over 252} P_{11}(k)\!\int\!dq  \,P_{11}(q)\,
\Biggl[ 12{k^2\over q^2}-158+100{q^2\over k^2}-42{q^4\over k^4}\nonumber\\
&&+{3 \over k^5 q^3} (q^2-k^2)^3 (7q^2+2k^2) \ln\left({k+q\over|k-q|}\right)
\Biggr]  \, .
\label{p13b}\end{eqnarray}

Thus with a specified initial spectrum $P_{11}(k)$ equations (17)
and (19) give the second order contribution. Before evaluating these
integrals for the CDM initial spectrum, we point out that the
poles of $F_2$ and $G_2$ described after equations (11)
give the leading order part of the integrand of equation (17) in $(q/k)$ as:
\begin{equation}
P_{22}(k)\sim k^2P_{11}(k)\int{d^3q\over3q^2}\,P_{11}(q)\ .
\label{p22div}\end{equation}
If $P_{11}(k)\sim k^n$ with $n\le-1$ as $k\to0$, then $P_{22}$ diverges.
Vishniac (1983) showed that the leading order part of $2 P_{13}$
in $(q/k)$ is negative and
exactly cancels that of $P_{22}$ --- this can be demonstrated by examining
the limiting form of $F_3^{(s)}$. In a future paper we will
analyze the leading order behavior of perturbative
integrals at higher orders and also calculate it using a nonperturbative
approach in order to investigate whether there may exist divergences for
some power spectra at higher orders in perturbation theory.
For the purposes of the second order integration the
cancellation of the leading order terms has no consequence other than
requiring that each piece, $P_{22}$ and $P_{13}$, be integrated
very accurately to get the resultant. This is necessary because the
cancelling parts cannot be removed before performing the integrals
as the two integrands have different forms:  $P_{22}$ is symmetric
in $\vec q$ and $(\vec k-\vec q)$, whereas $P_{13}$ is not. We will
return to this point in the next section.

\section{Results for CDM}

The results obtained in the previous section will now be used to
obtain the second order contributions to the CDM power spectrum.
We will use the standard CDM spectrum with parameters $\Omega=1$,
$H_0=50 \ {\rm km\, s^{-1}} {\rm Mpc^{-1}}$, and $\sigma_8=1$. For the
linear spectrum at $a=1$ we use the fitting form given by Bardeen et al.
(1986):
\begin{eqnarray}
&&P_{11}(k)=AkT^2(k)\ ,\quad A=2.19\times10^4 {\rm Mpc}^4\ ,\nonumber\\
&&T(k)={\ln(1+9.36k)\over 9.36k}\left[1+15.6k+(64.4k)^2+(21.8k)^3+
(26.8k)^4\right]^{-1/4}
\ ,
\label{pcdm}\end{eqnarray}
where $k$ is in units of Mpc$^{-1}$.
With this initial spectrum equations (17) and (19) can be used to obtain
the second order contribution $P_2(k)$, which can then be used to
obtain the net power spectrum as a function of $a$ and $k$ from equation (14).

\subsection{Nonlinear Power Spectrum }

As pointed out in Section 2.2 the integrals for $P_{22}$ and $P_{13}$
contain large contributions which exactly cancel each other.  For the CDM
spectrum these contributions are finite but care is still required in
their numerical evaluation.  Equal
contributions from $P_{22}$ are made as $\vec q \to0$ and $\vec q\to\vec k$,
whereas the cancelling contribution from $2 P_{13}$ is made only as $\vec
q\to 0$. The integrand for $P_{22}$ is symmetric in $\vec q$ and $(\vec k-
\vec q)$ and is positive definite.
For ease of numerical integration, we break up the integration
range for $P_{22}$ as follows:
\begin{eqnarray}
\int {d^3q\over 2\pi}&=&2 \int_0^\epsilon dq\int_{-1}^1
dy + \int_\epsilon^{k-\epsilon}dq\int_{-1}^1 dy +\int_{k-\epsilon}^
{k+\epsilon} dq \int_{-1}^{(k^2+q^2-\epsilon^2)/2kq} dy \nonumber\\
&&+\int_{k+\epsilon}^{k-k_c} dq\int_{-1}^1 dy +\int_{k-k_c}^{k_c} dq
\int_{(k^2+q^2-k_c^2)/2kq}^1 dy
\ ,
\label{limits}\end{eqnarray}
where $y\equiv \cos\vartheta$, and $k_c$ is the upper limit required
because at high $q$ the spectrum has departed strongly from the linear
spectrum causing the perturbative expansion to break down.  Transfer
of power from higher frequencies is suppressed by virialization.
The first term on the right-hand side of equation (22) has a factor of
$2$ because we have used the symmetry between $\vec q$ and $(\vec k-\vec
q)$ in the integrand to exclude a small ball of radius $\epsilon$ around
$\vec q=\vec k$ (where the integration becomes difficult)  by restricting
the limits on $y$ in the third term, requiring us to double the contribution
from a similar ball around $\vec q=0$ to compensate.  The limits on $y$
in the last term are set to ensure that $|\vec k-\vec q|\leq k_c$
as required to consistently impose the upper limit, i.e., to exclude any
contribution from $P_{11}$ in equation (17) when its argument exceeds $k_c$.
It is in principle important to scale $k_c$ with time to reflect the
growth of the nonlinear length scale with time,
because that determines the range
of validity of the perturbative expansion. We have done so using the
linear scaling $k_c\propto a^{-2/(3+n)}$, although as
explained below at early times the result is insensitive to the choice
of $k_c$.

The results of performing the integrals in equations (17) and (19) for
a large range of values of $k$ are shown in Figure 1.
We plot the linear spectrum $a^2 P_{11}(k)$, the
net spectrum including second order contributions given by equation (14),
and the nonlinear spectrum computed from high-resolution N-body simulations
described in Section 3.2
at four values of the expansion factor. The spectra have been divided
by $a^2$ to facilitate comparison of the results at different times.
The second order results at different
values of $a$  are obtained by simply multiplying $P_{11}$ and
$P_2$ by different powers of $a$ as shown in equation (14), so the
integration of $P_{22}$ and $P_{13}$ needs to be done only once for a
given $k$. The second order spectrum should be taken
seriously only for the range of $k$ for which
$a^4 P_2(k)<a^2 P_{11}(k)$, as we do not expect the perturbative
results to be valid for higher $k$.
The interesting range of $k$, the regime where nonlinear effects set in,
moves to lower $k$ as one looks at larger $a$, reflecting the progress
of nonlinearities to larger length scales (lower $k$) at late times.
As expected we find that at a given time
the second order contribution is not significant for small $k$
where the rms $\delta \rho / \rho \ll1$.

For small $k$ up to just over
the peak of the spectrum, the second order contribution is negative,
causing the nonlinear spectrum to be lower than the linear one. At
relatively high $k$ the second order contribution enhances the growth
of the spectrum.  This has the effect of making the slope of
the spectrum significantly shallower at high $k$ than that of the linear
spectrum. Thus, power is effectively transfered from
long to short wavelengths,
although the enhancement at short wavelengths exceeds the suppression
at long wavelengths.

The two power law model of Makino et al. (1992) gives qualitatively
similar results to those shown in Figure 1.
Bond \& Couchman (1988) also computed the
second order contributions to the CDM spectrum with a view to checking the
reliability of the Zel'dovich approximation at the same order.
They found excellent agreement,
in contrast to the results of Grinstein \& Wise (1987) who found that in
comparison to perturbation theory the
Zel'dovich approximation significantly underestimated the magnitudes of the
gaussian filtered, connected parts of the third and fourth moment of the
real space density. In comparison to our results, Figure 3 of Bond \& Couchman
shows a larger enhancement over the linear spectrum, and does not appear to
show the suppression at relatively low $k$ at all. They do not give the
explicit form of the term corresponding to our $P_{13}$, but state
that it is negligible in comparison to $P_{22}$. This does not agree with
our results at low $k$ and is probably the source of the difference in our
figures. It is difficult to make a more detailed comparison without knowing
the explicit form of their second term.

In order to obtain a  better understanding of the dynamics of the
mode-coupling, we have examined the relative contribution of different parts
of the CDM initial spectrum to the second order results at a given $k$.
Let $\vec q$ denote the integrated wavevector and
$\vec k$ the external wavevector at which the second order contribution is
calculated, as in equations (17) and (19).
There is a two-fold ambiguity because wavevector
$\vec k-\vec q$ contributes at the same time as $\vec q$.
We have carefully examined different ways of
associating second order contributions from different
parts of the initial spectrum, and found that the
second order contribution from $q\lsim k$ tends to be positive and that from
$q\gsim k$ negative.
Indeed we also find this to hold for power law spectra with
$-3\leq n\leq 1$, independently of the value of $n$, thus indicating that it
is a general feature of second-order mode-coupling.
In Figure 2 we associate the second-order contributions,
$dP_2(k)/d\ln q$,
with the smaller of $q$ and $\vert\vec k-\vec q\,\vert$.

There are two regimes in the CDM spectrum,
divided roughly by the part where the logarithmic slope $n \, [\equiv d\ln
P_{11})/d\ln k]$ falls below $-1$.
For small $k$, where $n\gsim -1$, the positive second order contribution from
$q\lsim k$ is swamped by the negative contribution made by large $q$.
The net effect is to decrease the growth of the spectrum compared
to the linear growth. For relatively large $k$, where $n\lsim -2$
the positive contribution from small $q$ dominates, increasingly so as one
goes to higher $k$. A comparison of the curves in Figure 2 for $k=0.1\,
{\rm Mpc^{-1}}$ and $k=1 \, {\rm Mpc^{-1}}$ shows how the relative strengths
of the
positive and negative contributions shift as one moves across the spectrum.
This shift can be understood by observing that at higher
$k$ there is an increasing amount of power in the initial spectrum at
$q<k$; the plot of the rms power on scale $k$ in Figure 3 illustrates
this point. The increased power at small $q$ causes a larger nonlinear
enhancement at
higher $k$. Since the weakly nonlinear regime moves to higher $k$ at
earlier times, the enhancement at high $k$ in turn leads to a stronger
nonlinear growth at
earlier times. We study the consequence of this fact in detail in Section 3.3.
The dominance of the nonlinear contribution from long-wave modes also
strengthens the consistency of the perturbative calculation,
because the amplitude of the density fluctuations is small for these
modes. As discussed in Section 3.2, this may be responsible for the second
order results being valid for a much larger range of scales at earlier times.

We emphasize that the transition value $n\simeq -1$ for the change in sign
of the second order contribution is only approximate, because it depends on
the value of $k$ taken as being representative of the weakly nonlinear
regime.
We have examined the second order contribution for power law
spectra $P_{11}(k)\propto k^n$, for a range of values of $n$
between $-3$ and $1$ to
verify this transition. We find that in the weakly nonlinear regime
(defined by $k\lsim k_{nl}$, where $k_{nl}$ is the scale at which
the rms $\delta \rho / \rho =1$), the second order contribution for
$n$ sufficiently larger than $-1$ is negative and that for $n$
sufficiently smaller than $-1$ is positive. For $n\simeq -1$, the
contribution is negative for low $k$ and positive for high $k$ in
the weakly nonlinear regime. These results are consistent with the
results of Makino et al. (1992) who examined the second order
contributions for $n=1,0,-1,-2$; they also found good agreement with
N-body simulations.

A possible transition in the nature of nonlinear evolution at $n=-1$
has also been explored by studying the clustering in real space in
N-body simulations by Klypin \& Melott (1992).
An examination of the origin of the term providing the dominant second
order enhancement suggests that the advective ($\vec v\cdot\vec\nabla $) terms
in the real space fluid equations cause the change in sign of the
nonlinear contribution. This interpretation is consistent with the fact
that for $n<-1$ there is an increasing amount of power in the rms
velocity field on larger scales, and this appears to cause the nonlinear
enhancement of the density from long-wave modes to dominate.
These arguments are by no means rigorous, and merit further exploration.

It is worth noting that for the deeply
nonlinear regime the stable clustering hypothesis (Peebles 1980,
Section 73) predicts that the spectrum
steepens below the linear theory spectrum for $n > -2$ and rises above
it for $n<-2$. Consistency with the second order results would require
that at least for $-2\lsim n\lsim -1$, the nonlinear spectrum first rise
above the
linear one in the weakly nonlinear regime and then fall below it in the
deeply nonlinear regime. This is indeed seen in
N-body simulations; the results of Efstathiou et al. (1988) show only hints
of this feature owing to limited resolution, but it is clearly evident in
simulations with higher resolution (Bertschinger \&
Gelb 1991; White 1993).

The second order spectrum provides an estimate of the change in the
fluctuation amplitude due to nonlinear effects in the weakly nonlinear
regime. The conventional normalization
is to set the rms $\delta \rho / \rho$
on a scale of $8 h^{-1}$ Mpc, denoted $\sigma_8$,
equal to $1$. The rms value is computed from the power spectrum using a
top-hat filter as described in Section 3.3. We find that with
the linear spectrum normalized in this way, second order effects increase
$\sigma_8$ by $10\%$. This is a smaller enhancement
than found by Hoffman (1987) for
the standard deviation of the density (without filtering) using the
Zel'dovich approximation. The N-body spectrum shows an even smaller change
in $\sigma_8$ than the second order spectrum, although it is difficult to
estimate accurately in a box of length 50 $h^{-1}$ Mpc.

\subsection{Comparison with N-Body Simulations}

The N-body results shown in Figure 1 are from two different
particle-particle/particle-mesh
simulations of the CDM model in a $(100\, {\rm Mpc})^3$ box normalized
so that linear $\sigma_8=1$ at $a=1/(1+z)=1.$  For $a>0.1$ we have
used the simulation with $144^3$ particles and Plummer softening distance
65 kpc performed by Gelb \& Bertschinger (1993).  To obtain accurate
results at higher redshifts we have performed a new simulation with
$288^3$ particles each of mass $2.9\times10^9\,M_\odot$ with Plummer
softening distance 20 kpc.  In both cases the energy conservation, as
measured by integrating the Layzer-Irvine equation, was much better than
1 percent.

The comparison of power spectra in Figure 1 shows qualitative
agreement between the second order and N-body results --- in both the small
dip in the spectrum at small $k$ and the enhancement at high $k$.
At early times the agreement of the two nonlinear spectra is
excellent.
This agreement extends beyond the naive regime of validity
of the second order results. As suggested above, the dominance of the
contribution from long-wave modes to the nonlinear enhancement at early
times apparently extends the regime of validity of the second order results.

At late times ($0.5\lsim a\lsim1$) the second order results at high $k$
show a larger enhancement of the spectrum in comparison to the N-body
results. There is a significant discrepancy in the two results even
within the expected regime of validity of the second order results.
This discrepancy, coupled with the good agreement at early times,
indicates that
even in principle the second order spectrum could not have agreed with the
N-Body spectrum shown in Figure 1 at all times. The simple dependence of
the second order spectrum on $a$ given in equation (14) is
incompatible with the dependence of the N-Body spectrum on $a$ for the
full range of $k$ lying in the nonlinear regime.

A part
of the discrepancy at late times could arise from the dependence of the second
order results on the upper cutoff imposed on the integrals. The cutoff
dependence is indeed the largest at late times: for $a\geq 0.5$ reasonable
variations in $k_c$ can change the result typically by over $10\%$.
Another source of disagreement could be that the N-body simulations
are done in a finite size box, therefore they have a small-$k$ cutoff.
Since the contribution from long-wave modes is positive, excluding these
modes could cause simulations to underestimate the nonlinear enhancement
of power. On comparing CDM simulations in boxes of sides $100$ and
$640$ Mpc we do find this to be true, but the difference is very small.
Thus neither of the two reasons
mentioned above explain the magnitude of the disagreement between the
second order and N-body spectra. A possible explanation is an inadequate
suppression of the second order spectrum due to collapse on small scales,
i.e., ``previrialization'' (Davis \& Peebles 1977; Peebles 1990). Indeed the
second order contribution from $q>k$ is negative, in
qualitative agreement with such a suppression,
but it should not be surprising if the magnitude of the suppression is
significantly underestimated.
Higher order perturbative contributions may well include some of this
suppression. Our analytic treatment neglects small-scale pressure and
vorticity, which should also suppress the nonlinear enhancement of power.
As we mention in Section 1, so far N-body studies designed to test this
hypothesis have concluded that small scale effects are negligible. However
these studies have not tested different initial spectra, and they have not
examined the power spectrum itself with as much dynamic range as our
simulations provide.

\subsection{Scaling in Time}

The nonlinear power spectrum can be used to construct statistical
measures of density fluctuations in real space. These can then
be used to study the most important consequence of the
coupling of long-wave modes: a systematic change in the variation
of characteristic nonlinear scales with time. We proceed to
do this by first defining the rms $\delta\rho/\rho$
averaged on length scale $R$ by integrating over the power spectrum
with an appropriate window function $W$:
\begin{equation}
\delta_R^2(a)\equiv
\left\langle{\left(\delta\rho\over\rho\right)^2}\right\rangle_R=\int
d^3k \, P(a,k) \, W^2(kR)\ .
\label{drho}\end{equation}
For $W$ we shall use three different functions: a shell in $k$-space,
the top-hat in real space, and the gaussian, given respectively
by,
\begin{mathletters}
\begin{equation}W^2_{\rm D}(kR)=\delta_{\rm D}(kR-1)\ ,
\label{wd}\end{equation}
\begin{equation}W_{\rm TH}(kR)={3\,[\sin(kR)-kR\cos(kR)] \over (kR)^3}\ ,
\label{wth}\end{equation}
\begin{equation}
W_G(kR)=\exp\left[-{(kR)^2\over 2}\right] \ .
\label{wg}\end{equation}
\end{mathletters}

In Figure 3 we plot $[4\pi k^3P(a,k)]^{1/2}$, or $\delta_R(a)$
for $W_{\rm D}$ with $k=r^{-1}$, to illustrate what we expect
for the time dependence of a characteristic nonlinear scale, denoted as
$R_{\rm nl}(a)$.
If the spectrum evolved self-similarly then one would expect
that at all $a$, the onset of nonlinear effects occurs at a scale
defined by setting
$4\pi k^3P(a,k)=\hbox{constant}$ for some value of the
$\hbox{constant}$ of order unity. This behavior is expected for power law
spectra of the form $P(k)\propto k^n$, and has been verified in studies of
N-body simulations (Efstathiou et al. 1988).
Even though CDM-like spectra are not pure power laws, the simplest assumption
would be that they show a similar behavior. However, Figure 3 shows that
at early times (small $a$), the spectrum deviates from the linear one at
progressively smaller values of $4\pi k^3P(a,k)$. This trend is even stronger
for the N-body spectra. Thus already there is a hint
of a systematic departure of the nonlinear scaling from the conventional
expectation, due to the variation of the spectral index
$n$ with scale for the CDM spectrum.

This conclusion is confirmed by using the other window functions to define
the nonlinear scale as follows.   We calculate
$\delta_R(a)$ using equations (24b,c) for a range of $R$ and $a$.
We then define $R_{\rm nl}(a)$ as follows:
\begin{equation}
\delta_R
(a)=\delta_c\quad\hbox{for}\quad R=R_{\rm nl}(a)\ ,
\label{rnl}\end{equation}
where $\delta_c$ is a constant of order unity.
Since $R_{\rm nl}(a)$ is a comoving length scale, it can be used
to define a nonlinear scale for the mass as:
$M_{\rm nl}(a)=(4\pi/3)\bar \rho R^3_{\rm nl}(a)$, where
$\bar \rho$ is the critical density today.

In Figure 4 we have plotted
$M_{\rm nl}(a)$ from $a=0.04$ to $1$ for the gaussian and top-hat filters,
with $\delta_c$ chosen to be $1$ and $1.69$ for each filter.
The dependence of $M_{\rm nl}(a)$
confirms the impression conveyed by Figure 3: nonlinear enhancement
is stronger at earlier times. While the quantitative results depend on
the choice of the window function and $\delta_c$, it is clear that in
each of the figures the slope of the second order curve is different
from the linear curve, and this causes the relative enhancement of
$M_{\rm nl}(a)$ to be larger at earlier times. Indeed, if the normalization
of the second order curves was changed (thus shifting them to the right) so
that at $a=1$ they predicted the same nonlinear mass as the linear curves,
then all four panels would show very similar relative enhancements at early
times.

In stating quantitative results for the time-dependence of nonlinear
masses we shall focus on the
gaussian filter with $\delta_c=1$. This choice provides the most
conservative estimates of second order effects.
At $\, a^{-1}=(1+z)= (20,10,5,2)$, $M_{\rm nl}(a)$ from the second order
spectrum is about $(180,8,2.5,1.6)$ times (respectively)
larger than the linear case. Figure 4 can also be used to
read off the change in the redshift of nonlinearity for the desired mass
scale due to second order effects.
(Here as in the preceding figures, the linear spectrum is normalized so
that $\sigma_8=1$ at $z=0$ and this fixes the normalization of the
second-order spectrum.) For example, the mass scale $10^6 \,M_\odot$
goes nonlinear at $(1+z)\simeq 25$ as opposed to 19 if only the
linear spectrum is used; and the mass scale $10^{11} \,M_\odot$ at $(1+z)
\simeq 6$ as opposed to 5. This change in redshift is a more
meaningful indicator of the nonlinear effect, as the change
in $M_{\rm nl}(a)$ is amplified due to the steepening of the spectrum at
high $k$.

In Figure 4 we have also shown results from
the Press-Schechter model (Press \& Schechter 1974, hereafter PS).
The PS model is a widely used ansatz for predicting the distribution of bound
objects of a given mass at different times (section 3.4). It relies on the
linear growth of the power spectrum, hence it is no surprise that the shape
of the PS curve is very similar to the linear curve. Here the PS nonlinear
mass is defined as the mass for which a fixed fraction, $0.4$, of the
mass in nonlinear clumps belongs to clumps of mass $M_{\rm nl}$ or larger.
The fraction $0.4$ is chosen so that the normalization of the PS
curves is close to that of the other curves at $a=1$ --- in the upper
panels it is close to the nonlinear curves and in the lower panels to the
linear curves.

The N-body simulations can be used to define a characteristic nonlinear
mass in many different ways. The dashed curves in Figure 4 show
nonlinear masses computed using the power spectrum from the simulation
in the same way as for the second order and linear spectra above (i.e.,
using eq. \ref{rnl}). The results
are in very good agreement with the second order results, as expected
because of the good agreement of the second order and nonlinear power
spectra.  The relative enhancement of $M_{\rm nl}(a)$ over the linear
prediction at $(1+z)= (10,5,2)$ is $(11,2.4,1.2)$.
By examining all four panels it can be seen that, independent of the parameters
used, the slopes of the curves using second-order and nonlinear power
spectra are distinctly different
from those of the linear and PS curves.  The filled triangles
use a different definition of the N-body characteristic nonlinear mass and
will be discussed in the next section.

Our results in Figure 4 indicate that linear scaling for $M_{\rm nl}(a)$
significantly
underestimates nonlinear enhancement at high redshift. Consequently the
characteristic masses predicted by the PS model are much smaller
than the second order and N-body masses for $z>4$, even for the choice
of parameters for which they agree at late times. This conclusion may
appear at odds with previous tests of the PS formalism made by others.
However, no previous tests have examined the CDM model at high redshift
with as much dynamic range as we have.  As we have emphasized, mode-coupling
from long waves is strongest for small $n$; for the CDM spectrum $n$ varies
with scale and approaches $-3$ at high $k$.  It is precisely in this limit,
previously untested with high-resolution N-body simulations, that we find
the greatest departures from linear theory and the Press-Schechter model.

\subsection{Distribution of Nonlinear Masses}

The characteristic nonlinear masses defined above do not fully characterize
the distribution of dense clumps that form as a result of gravitational
instability.  A better comparison of theory and simulation can be made
using the complete distribution of masses.

In the N-body simulation we have identified dense clumps at mean
overdensity about 200 using the friends-of-friends (FOF)
algorithm with linking distance 0.2 times the mean interparticle separation.
(The $N=288^3$ simulation was used at $1+z=10$, while the
$N=144^3$ simulation was used for $1+z=5$, 2, and 1.)
The distribution of nonlinear clump masses is very broad, so there is no
unique nonlinear mass.  We have chosen to define the characteristic nonlinear
mass $M_{\rm nl}$ for this distribution as the median clump mass defined
so that half of the mass in clumps of at least 5 particles is contained in
clumps more massive than $M_{\rm nl}$.  The 5 particle limit corresponds
to $M=1.16\times10^{11}\,M_\odot$ and $1.45\times10^{10}\,M_\odot$ for
the $N=144^3$ and $288^3$ simulations, respectively.  The resulting nonlinear
masses are denoted by the filled triangular symbols in the upper-left
panel of Figure 4; for other panels these points would be at the same
locations as in this panel.  (If the PS curves were defined with the same
lower limit for clump masses and the same value of the mass fraction,
instead of having no lower limit and a mass fraction 0.4, they would
agree more closely with the N-body FOF points.) It is coincidental that
this definition of N-body nonlinear mass yields such close agreement with the
analytic predictions at $z=0$, because the broad range of clump masses
would allow us to vary $M_{\rm nl}$ by factors of a few.  The relative
variations as a function of redshift are more meaningful. It is clear from
Figure 4 that the variation of these N-body masses with $a$ departs from
the linear scaling even more strongly than the curve computed from
the N-body power spectrum.  Thus, nonlinear effects on the formation of
high-redshift objects appear to be even more significant than they are on the
power spectrum.  However, the 5 particle limit affects the FOF characteristic
mass (no lower limit is imposed on the PS curve), so we should make a
more detailed comparison with the mass distribution before reaching firm
conclusions.

The PS model makes the ansatz that the formation of bound objects
is determined by the overdensity in the linear density field. Using the
gaussian distribution of the linear density field, this ansatz gives the
comoving number density, $n(M,a)$, of nonlinear objects of mass $M$ in the mass
interval $dM$ at expansion factor $a$ as (Press \& Schechter 1974):
\begin{equation}
n(M,a)\, dM\, =\bar\rho\left(2\over\pi\right)^{1/2}{\delta_c\over
\sigma}\, {\rm exp}{\left(-{\delta_c^2\over {2 \sigma^2}}\right)}{1\over\sigma}
\left({d\sigma\over dM}\right){dM\over M}\, .
\label{ps}
\end{equation}
In this equation $\delta_c$ is a free parameter which can be taken to
be a constant, with the linear rms density smoothed on the mass
scale $M$, $\sigma(M,a)$, growing in proportion to $a$.
A popular choice for $\delta_c$ is $1.69$, the value of the linear
density at which a spherical top-hat perturbation collapses to infinite
density.
The PS mass distribution $n(M,a)$ has been tested against N-body simulations
and found by other workers to work very well. Efstathiou
et al. (1988) tested it for scale-free simulations, and several authors
have tested it for the CDM spectrum (e.g., Carlberg \& Couchman 1989).
The weaknesses of such tests --- particularly, the finite resolution of
the simulations --- have been recognized by these authors, but even
so the agreement has been surprisingly good for the range of masses and
redshift probed. Consequently, the PS model has been
widely used in predicting the number density of objects at high redshift, or
in estimating the redshift at which a given mass scale goes nonlinear.

Figure 5 shows the cumulative mass fraction (CMF) as a function of clump
mass from the N-body simulation and the PS prediction. The CMF is defined
by
\begin{equation}
\hbox{CMF}(M,a)=\bar\rho^{-1}\int_M^\infty n(M,a)M\,dM\ ,
\label{cmf}
\end{equation}
i.e., the fraction of mass in objects of mass $M$ or larger.
In Figure 4 we defined the PS nonlinear mass using the condition
$\hbox{CMF}=0.4$.
For the PS prediction of Figure 5 we have chosen the top-hat filter with
$\delta_c=1.69$.  At late times, this choice gives fairly good agreement
for the high-mass end of the mass distribution.  However, we see that at
early times the N-body mass distribution lies systematically above the PS
prediction.  This is in qualitative agreement with the results shown in
Figure 4 and supports our conclusion that nonlinear effects on the formation
of nonlinear clumps are even stronger than they are for the power spectrum.

Because coupling of long waves modifies the power spectrum and therefore
the rms density $\sigma(M,a)$, the failure of Press-Schecter theory to
match the N-body results exactly does not surprise us.  As an experiment
we replaced $\sigma(M,a)$ in equation (\ref{ps}) using the second-order and
N-body power spectra instead of linear theory.  The resulting $\hbox{CMF}(M)$
falls too rapidly at large $M$, even after $\delta_c$ is increased to
compensate for the nonlinear enhancement of density fluctuations.  If
the nonlinear power spectrum is used the PS formula gives the wrong
shape for $n(M,a)$ because it assumes a gaussian distribution of densities,
while the nonlinear density field has a broader distribution.  We have
found no simple modification of the PS formula that can account for the
systematic departures evident in Figure 5.  Expressing an optimistic view,
we note that the PS formula is accurate to about a factor of 2 for the CMF
over the entire range shown in Figure 5.  On the other hand, the deviations
are larger for rarer objects (smaller CMF) and the sign and magnitude of
the deviation changes systematically with $a$.  Therefore one should use
the PS formula, especially at high redshift and for rare objects, only
with caution after calibration by high-resolution N-body simulations.

\section{Discussion}

We have calculated the second order contribution to the evolution of
the standard CDM power spectrum. We believe that our results capture the
dominant nonlinear contribution in the weakly nonlinear regime.
They are consistent with N-body results in this regime from $z=9$ to
$z\simeq 1$, but show a larger enhancement of the spectrum than the N-body
results from $z\simeq 1$ to $z=0$. The bulk of the second order enhancement in
the growth of the power spectrum is provided by the mode coupling of
long-wave modes, especially for the onset of nonlinearities at high redshift.

By analyzing the perturbative integrals we have studied the sensitivity of
nonlinear evolution to different parts of the spectrum, and thus have
probed the dynamics of the mode-coupling at work. We find that on
scales of interest to large-scale structure in the universe,
the dominant contribution to the weakly nonlinear evolution of most
realistic power spectra comes from the mode-coupling
of long-wave modes. Perturbation theory is quite adequate for
estimating this contribution since the amplitude of density fluctuations
is small for the long-wave modes.

An important consequence of nonlinear evolution is to
change the time dependence of the nonlinear scale $M_{\rm nl}(a)$ from
linear scaling: it is found to be significantly
larger at high $z$. Thus objects of a given scale go nonlinear at
higher redshifts than indicated by the standard linear extrapolation.
As discussed in Section 3, this is a consequence of the variation
with scale of the spectral index, with $n\gsim -1$ on the scales of
interest for large-scale structure and $n\simeq -3$ on the smallest scales.
We have given quantitative estimates of this effect for the standard CDM
spectrum for different window functions and definitions of nonlinear scale.
For a gaussian window function and $\delta_c=1$, which provides the most
conservative estimates, the
change in the redshift factor of nonlinearity, $(1+z_{nl})$,
is about $20 \%$ for $10^{11} \,M_\odot$ objects (with linear extrapolation
$1+z_{nl}=5$) and increases to about $33 \%$ for $10^{6} \,M_\odot$ objects
($1+z_{nl}=19$). We have also computed nonlinear corrections using high
resolution N-body simulations, using the power spectrum from the simulations
as well as directly identifying bound objects.
The results are in very good agreement with
the second order predictions, especially between $z\simeq 4$ and $10$.
Quantitative comparisons are provided in Sections 3.3 and 3.4 and in
Figure 4.

Thus the most striking implications of second order effects
are for the formation of nonlinear objects at high $z$.
Theoretical studies of, for example,
the first generation of collapsed objects,
the redshift of galaxy-formation, and reionization at high-$z$ (see e.g.,
Couchman \& Rees 1986; Efstathiou \& Rees 1988; Tegmark \& Silk 1993) ---
all require as an input the scale of nonlinearity as a function of $z$.
For analytical estimates this is invariably obtained using linear
extrapolation, as for example in the Press-Schechter mass distribution.
We have shown (Figure 5) that the Press-Schechter theory leads to a systematic
underestimate of the abundance of high-mass nonlinear clumps at high
redshift in the CDM model, but have not succeeded in suggesting a
simple modification that works better.  While nonlinear coupling to
long waves increases the amplitude of small-scale density fluctuations,
it also changes the probability distribution from the gaussian distribution
appropriate in the linear regime.

Most realistic cosmological spectra steepen to $n \simeq -3$ at the smallest
scales and have $n\gsim -1 $ on the largest scales of interest.
This is a generic
feature arising from the sluggish, logarithmic growth of fluctuations during
the radiation dominated era, thus causing the scale invariant spectrum with
spectral index $n=1$ initially to approach $n=-3$ on the smallest
scales while retaining the primeval slope
on scales much larger than the
size of the horizon at the end of the radiation dominated era.
Hence for different cosmological
models the basic features of nonlinear gravitational evolution
that we have studied should hold, although the
quantitative details would depend on the
values of parameters such as $\sigma_8$, $\delta_c$, $\Omega$ and $H_0$.

The increase in redshift of collapse relative to linear theory
that we have calculated for CDM
should also occur in all realistic spectra provided that on the scales
of interest $n$ decreases sufficiently rapidly with increasing $k$.
Our results will not
apply if the dark matter is hot, but the qualitative implications should
be the same for the evolution of the baryonic component in a CDM- or
baryon-dominated model until dissipational effects become important.
For spectra with a very steep slope at small scales
(such as in the hot dark matter model),
second order effects may lead to a strong
nonlinear enhancement which would drive the spectrum to a shallower
slope.

In the near future second order power spectra from theoretical models
could be related to the power spectrum
calculated from observational surveys. Indeed the shape of the
best fit three-dimensional power spectrum computed from results of the APM
survey (Baugh \& Efstathiou 1993) shows two characteristic features of
the second order CDM spectrum: a relatively shallow slope at small scales
and a flattening of the peak of the spectrum at large scales.
The power spectrum computed from the CfA redshift survey (Vogeley et al.
1992) and from the $1.2$Jy IRAS redshift survey (Fisher et al. 1993)
had also shown the first feature of a shallow slope with $n$ just below $-1$
at high $k$,
but these surveys lacked the depth required to determine the shape of
the spectrum near the peak. It will be interesting to see if the extended
peak of the APM spectrum is a robust feature.

\acknowledgements

It is a pleasure to thank Alan Guth for many stimulating discussions.
We also acknowledge useful discussions with Carlton Baugh,
George Efstathiou, Yehuda Hoffman, Roman Juszkiewicz, David Weinberg
and Simon White.  We thank John Bahcall for his hospitality at the
Institute for Advanced Study, where this work was completed.
Supercomputing time was provided by the Cornell National Supercomputer
Facility and the National Center for Supercomputing Applications.
This work was supported by NSF grant AST90-01762.

\clearpage

\section{Figure Captions}

\noindent Fig. 1:
Linear and nonlinear power spectra at expansion factors $a = 0.1, 0.2, 0.5,$
and $1$, where $a=1$ corresponds to linear $\sigma_8=1$.  The linear spectrum
is given by the dotted curves, the corresponding second order spectrum
[$P(k)=a^2P_{11}(k)+a^4P_2(k)$] by the solid curves
and the spectrum from high resolution N-body simulations by the
dashed curve. The spectra are all divided by $a^2$ to facilitate
better comparison of the nonlinear effects at different values of $a$.
The triangles marked on the second order spectra indicate the point
at which $a^4P_2(k)=a^2P_{11}(k)$: this indicates the
approximate limit of validity of the second order results.

\noindent Fig. 2:
Contributions to $P_2(k)$ vs. $\log q$, where $q$ is the magnitude of
the integrated wavevector.
The two panels are for different choices of $k$.  $P_2(k)$ is defined in
equation (15) and is the sum of the contributions $P_{22}(k)$ and
$2 P_{13}(k)$. The integrand of $P_{22}(k)$ is symmetric in $\vec q$ and
$(\vec k - \vec q )$; we have chosen to associate the contribution from
such a pair of wavevectors with the wavevector with smaller magnitude.
Other choices do not alter the basic trend seen here, namely, that the
contribution from $q<k$ is generally positive and peaked at $q=k/2$,
while that from $q>k$ is generally negative. Moreover, a comparison of
the plots for the two values of $k$ shown illustrates
that at higher $k$ the positive contribution from small $q$
dominates, leading to a net enhancement of small-scale power.
This is due to the increasing amount of power at $q<k$ for
higher $k$, as can be seen in Figure 3.

\noindent Fig. 3:
RMS amplitude of density fluctuations
vs. scale $k$ for several expansion factors.
The lower curves correspond to smaller $a$.
Solid (dashed) curves are used for the second order (linear) results.
The second order curves are shown only for the estimated regime of
validity shown in Figure 1. It is clear from the results at different
$a$ that the nonlinear contribution becomes significant at earlier $a$
for successively smaller values of $[4\pi k^3 P(a,k)]^{1/2}$.

\noindent Fig. 4:
Growth of characteristic nonlinear mass with time.
The mass scale $M_{\rm nl}(a)$ at which the rms $\delta\rho/\rho$
reaches a fixed value (denoted by $\delta_c$ in the figures) is plotted
vs. the expansion factor $a=1/(1+z)$.  Two values of $\delta_c$ are used
to define $M_{\rm nl}(a)$: $\delta_c=1$ on the left and $\delta_c=1.69$
on the right. For each $\delta_c$ the rms $\delta\rho/\rho$ is computed
with a gaussian window function for the upper panels, and with a real
space top-hat for the lower panels. The dotted curves show $M_{\rm nl}(a)$
computed using the linear spectrum $P_{11}(k)$; the solid curves include
the second order contribution for the same normalization of the linear
spectrum. The dot-dashed curves have been computed from the
N-body power spectrum shown in Figure 1. The dashed curves are
computed using the Press-Schechter model, with the characteristic
nonlinear mass defined as
that at which a fixed fraction, $0.4$, of the mass in nonlinear clumps
is in clumps more massive than $M_{\rm nl}$.
In the
top-left panel the symbols labelled ``N-body FOF'' are obtained from the
N-body simulation by using the friends-of-friends algorithm to identify
clumps of at least 5 particles, and then to define a characteristic mass
so that half the mass in clumps is in clumps more massive than $M_{\rm nl}$.

\noindent Fig. 5:
Cumulative mass fraction (CMF) vs. clump mass $M$ at $a=0.1,0.2,0.5,1$.
The dashed curves represent the predictions of the Press-Schechter model,
while the solid curves are obtained from N-body simulations. The curves
are shown at different times, with the higher curves representing larger
values of $a$. The N-body curves are obtained using the friends-of-friends
algorithm with linking parameter $=0.2$.

\end{document}